\documentclass{pj}
\usepackage{graphicx}
\usepackage{cite}
\usepackage[update,prepend]{epstopdf}

\begin{document}
\setcounter{page}{1}
\pjheader{Vol.\ x, y--z, 2012}

\title[Magneto-Optical Effects Enhancement...]
{Magneto-Optical Effects Enhancement in DMS Layers Utilizing 1-D Photonic Crystal}
\author{M.~Koba}

\address{
Institute of Experimental Physics, Faculty of Physics, \\ University of Warsaw
\\ ul. Ho\.za 69, PL-00-681 Warszawa, Poland}

\address{
National Institute of Telecommunication
\\ul. Szachowa 1, PL-04-894 Warszawa, Poland}

\author{J.~Suf\mbox{}fczy\'{n}ski}

\address{
Institute of Experimental Physics, Faculty of Physics, \\ University of Warsaw
\\ ul. Ho\.za 69, PL-00-681 Warszawa, Poland}

\runningauthor{Koba}
\tocauthor{M.~Koba}

\begin{abstract}
We propose a theoretical design of a structure combining a diluted magnetic semiconductor (DMS) layer with one-dimensional photonic crystal and show that it allows for a sizable (a few-fold) enhancement of the magneto-optical Kerr effect (MOKE) magnitude. We base our calculations on the transfer matrix
method assuming realistic, and optimal with respect to a sample growth, parameters. We investigate representative cases of (Ga,Mn)N or (Cd,Mn)Te DMS layers deposited on the top of a respectively GaN- or CdTe-based distributed Bragg reflector, and highlight the applicability of the proposed design for both: III-V wurtzite and II-VI zinc-blende DMS.
\end{abstract}

%

\section{Introduction}

Diluted Magnetic Semiconductors (DMSs) are semiconductors in which a fraction of the cations is substituted with transition metal ions, e.g., Mn or Fe. Different types of DMSs comprising III-V, II-VI, and IV-VI materials exhibit a diversity of unique magnetic, optical and conductive properties attractive from the point of view of fundamental research as well as possible applications. The $s,p-d$ interaction between band carriers and localized magnetic moments in DMS leads to such effects as giant Zeeman splitting of bands or giant magneto-resistance, both beneficial for the domain of information processing and storage. GaN based DMSs are particularly appealing in the view of possible practical implementations, as they are expected to exhibit a ferromagnetic ordering of spins with a critical Curie temperature exceeding room temperature \cite{Dietl-Science-2000}.

As it was shown \cite{Ando-apl82-2003, Ando-science-2006} a correlation of magneto-optical and magnetic properties is an essential condition for the unequivocal assessment of the intrinsic nature of magnetization of DMS. The magneto-optical Kerr effect (MOKE) \cite{Kerr} is a commonly used optical tool for a determination of the DMS magnetization. Its magnitude scales typically linearly with the Zeeman splitting of excitonic levels which is in turn linearly proportional to the sample magnetization \cite{Gaj:1978,Gaj:1979}. However, the magneto-optical signal from DMS decreases at a very low and a very high magnetic ions doping level. In the first case it is due to a low magnitude of the Zeeman splitting, while in the latter one, due to the decrease of the optical signal resulting from an enhancement of a non-radiative carrier recombination related to ions.\cite{Lee2005} This limits the efficacy of MOKE, and other optical methods in DMS magnetization determination. To the contrary, optical effects in semiconductor may be enhanced by implementation of photonic crystals (PC). The PC that are periodic dielectric nanostructures containing regions of alternate high and low refractive index. They affect the propagation of the electromagnetic wave similarly as the periodic potential in a crystal affects the motion of electrons and exhibit, in particular, a forbidden gap in frequency domain.\cite{mit:abinitio}
Although it was highlighted in a number of works \cite{Gourdon-solidstatecomm123-2002,Liu-APL90-2007,Shimizu,Brunetti-PRB73-2006,Qureshi, Sadowski1987} that combining the advantages offered by both domains: DMS and photonics, may enforce a new capabilities in current and future devices, until recently DMS and photonics paved their way in the modern physics mostly separately.

Our work demonstrates that implementation of even a simple photonic structure enables a meaningful enhancement of magneto-optical effects in DMSs. We present a theoretical design of a DMS layer deposited on the top of a single distributed Bragg reflector (DBR) and show that such design leads to at least a few-fold enhancement of the MOKE magnitude. Our approach is different from the one developed in, e.g., Refs. \cite{Gourdon-solidstatecomm123-2002, Liu-APL90-2007, Shimizu,Brunetti-PRB73-2006}, where the enhancement of magneto-optical Faraday effect was obtained by enclosing a DMS layer within a photonic cavity made of two DBR mirrors, or e.g., Ref. \cite{Qureshi}, where in order to enhance the MOKE, a layer with single-domain nanomagnets was placed under a dielectric multilayer. The simplicity of the proposed design opens a way for its wide practical implementation. It is worth emphasizing that, with respective material parameters, the calculations can be easily applied to any stratified structure and essentially to any III-V and II-VI DMS material with any magnetic ion doping different from Mn, e.g., Fe or Co.

In what follows we introduce the structure design (Sec. \ref{sec:structure}) and a numerical model upon which we base our calculations (Sec. \ref{sec:model}), we present and discuss the  results (Sec. \ref{sec:numerical_results}), and finally conclude our work (Sec. \ref{sec:summary}).

\section{Structures}
\label{sec:structure}

Our investigation is focused on, but not bound to, the representative cases of Mn-doped wurtzite III-V and zinc-blende II-VI structures, that is (Ga,Mn)N and (Cd,Mn)Te, respectively. We consider realistically designed structures. We assume a presence of a buffer layer, as it is often introduced to GaN based structures grown on sapphire substrates in order to reduce density of dislocations limiting the optical signal quality. Moreover, we sustain a small difference of refractive indices of layers constituting the mirror, thus small contrast of their compositions, in order to limit the undesirable effects of strain and dislocation density resulting from the lattice mismatch.

\begin{figure}[h]
\centerline{\includegraphics[width=0.8\columnwidth,draft=false]{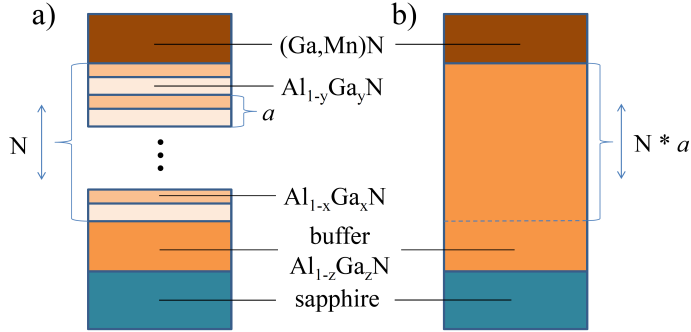}}
\caption{
a) The proposed sample design: DMS layer deposited on DBR mirror, grown on buffer deposited on a sapphire substrate. In the case of (Ga,Mn)N sample where $a$ = 61 nm, and the number of periods N varies from 1 to 15. b) Reference sample: DBR has been replaced by Al$_{0.1}$Ga$_{0.9}$N buffer layer of the thickness $Na$ equal to the one of DBR mirror.
}
\label{fig:structure_schemes}
\end{figure}

We perform our calculations for two structures shown schematically in Fig. \ref{fig:structure_schemes}: one representing the proposed design, involving DMS and DBR layers (Fig. \ref{fig:structure_schemes}a)), and the other one serving as a reference (Fig. \ref{fig:structure_schemes}b)) being just the DMS layer deposited on the buffer layer (no DBR).

More specifically, in the case of GaN-based structure, the design presented in Fig.~\ref{fig:structure_schemes}a) consists of the (Ga,Mn)N layer (x$_{Mn}$=0.06\%) thick for $d_{DMS}=100$~nm, DBR layer constituted by $N = 1$ to $15$ periods (thickness $a=61$ nm) of alternating Al$_{0.05}$Ga$_{0.95}$N/Al$_{0.2}$Ga$_{0.8}$N layers (with thicknesses of 28 nm and 33 nm, respectively), and Al$_{0.1}$Ga$_{0.9}$N buffer (thickness $d_{{buf\!fer}} = 200$~nm) deposited on a sapphire substrate. The thicknesses of the DBR mirror layers are tuned so the mirror stop-band spectral position is matched to near the band gap spectral region. The assumed material parameters are based on values taken from experiment \cite{Suffczynski:PRB11} to assure that the result of our numerical calculation is as close to real structure performance as possible. 
In the reference structure shown in Fig.~\ref{fig:structure_schemes}b) the DBR layer is replaced by the Al$_{0.1}$Ga$_{0.9}$N layer of the corresponding thickness $Na$. The structure parameters for CdTe based sample are given in Sec.~\ref{sec:numerical_results}.

\section {Model}
\label{sec:model}

The Magneto Optical Kerr Effect (MOKE) is a consequence of the magnetic circular birefringence of the medium: the linearly polarized electromagnetic wave after reflection from the surface of the sample magnetized parallel to the light propagation becomes elliptically polarized. This occurs due to magnetic field induced splitting of bands and resulting difference of refraction coefficient for two opposite circular polarizations of the light.
The angle between the major axis of the elliptically polarized reflected light and the polarization direction of the incident light defines the Kerr rotation 
$\Theta_K$, which is given by (e.g., Ref. \cite{Testelin}):

\begin{equation}
\Theta_K=\frac{1}{2}arg\frac{r^-}{r^+},
\label{eq:Theta_K}
\end{equation}
where ${r^-}$ and ${r^+}$ represent complex amplitudes of reflectivity coefficients for ${\sigma^-}$ and ${\sigma^+}$ circular polarizations, respectively.

In order to obtain both ${r^-}$ and ${r^+}$ values for an arbitrary one-dimensional stack of layers we perform calculations basing on the transfer matrix method (TMM) (e.g., Ref. \cite{Yeh}) taking into account contributions from excitons, their excited states, and unbound states to the dielectric functions of respective layers.

\begin{figure}[h]
\centerline{\includegraphics[width=0.8\columnwidth,draft=false]{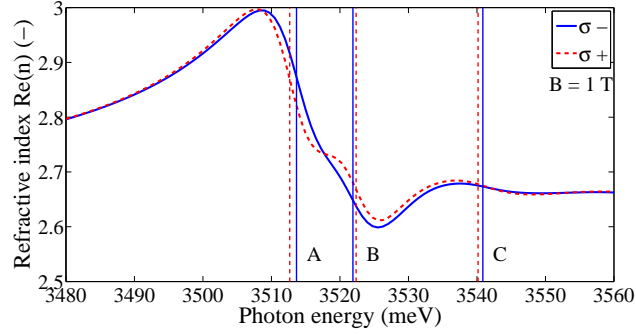}}
\caption{
Calculated real part of dielectric function of Ga$_{0.994}$Mn$_{0.006}$N at B = 1 T and T = 2 K for $\sigma-$ (solid line) and $\sigma+$ (dashed line) polarizations of the light. Vertical lines indicate respective A, B, and C exciton energies.
}
\label{fig:diel_fun_GaMnN}
\end{figure}

The calculations are made for two circular polarizations of the light and the case of external magnetic field of B = 1 T. The determination of exciton energies is based on a model involving effective excitonic Hamiltonian of the following form \cite{Suffczynski:PRB11,Pacuski:PRL2008,
Julier:PRB1998,Stepniewski:PRB1997,GilBriot}:

\begin{equation}
H = {E_0} + {H_{vb}} + {H_{e-h}} + {H_{diam}} + {H_Z} + {H_{s,p-d}}.
\label{eq:hamiltonian_general}
\end{equation}

In the above equation $E_0$ denotes the (Ga,Mn)N band-gap energy, $H_{vb}$ describes the valence band at $k$ = 0 in a wurtzite semiconductor, $H_{e-h}$ relates to the e-h exchange interactions within the exciton, $H_{diam}$ corresponds to an excitonic diamagnetic shift, $H_Z$ is an ordinary Zeeman excitonic Hamiltonian, and $H_{s,p-d}$ accounts for $s,p-d$ exchange interaction between excitons and localized Mn ion spins.

Parameters of excitonic transitions and of the electronic band structure in magnetic field are taken from Ref.~\cite{Suffczynski:PRB11} (or Ref.~\cite{Gaj:94} in the case of II-VI structure, considered below).
The real part of the dielectric function representing the refractive index of (Ga,Mn)N at B = 1 T for ${\sigma^-}$ and ${\sigma^+}$ circular polarizations of the light is depicted in Fig. \ref{fig:diel_fun_GaMnN}.
The contribution to a dielectric function from A, B and C excitons is seen. The excitons split in magnetic field, as indicated by vertical lines. Splitting of excitons A and B occurs toward opposite directions. It is also seen that the real parts of dielectric function for ${\sigma^-}$ and ${\sigma^+}$ polarizations differ only in the excitonic region. It will be shown in the next section that the exciton splitting has a dominant contribution to MOKE in the near the band gap region.

\section{Numerical results}\label{sec:numerical_results}
\begin{figure}
\centering
\centerline{\includegraphics[width=0.8\columnwidth,draft=false]{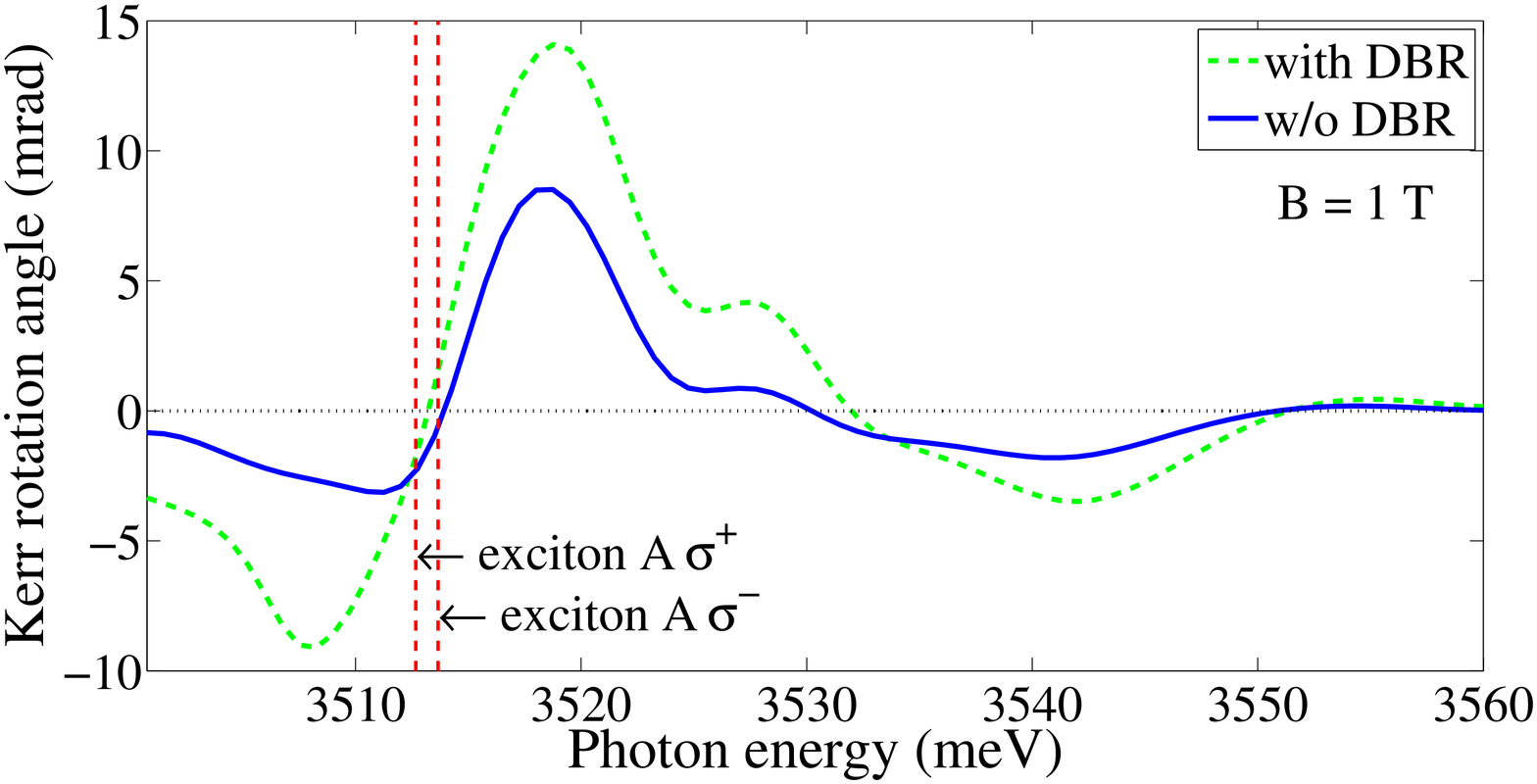}}
\caption{
Calculated Kerr rotation angle versus photon energy at B = 1 T for a structure containing (Ga,Mn)N layer with (dashed line) and without (solid line) DBR underneath, as described in the text. The energy position of the exciton A is indicated for both circular polarizations of the light for a reference.
}
\label{fig:Kerr_PE}
\end{figure}

Following the solution of equation \ref{eq:Theta_K} for parameters and dielectric functions given in Section \ref{sec:structure} and references cited there, we plot Kerr rotation angle as a function of photon energy, as presented in Fig. \ref{fig:Kerr_PE}.

Fig. \ref{fig:Kerr_PE} shows that the magnitude of the MOKE increases thanks to the implementation of the proposed structure design. Comparison of the peak-to-peak as well as area under the curve gives evidence for around two times the MOKE enhancement. Furthermore, for some photon energy values (e.g., 3506 - 3511 meV) the observed enhancement is as large as four.

\begin{figure}[h]
\centering
\centerline{\includegraphics[width=0.8\columnwidth,draft=false]{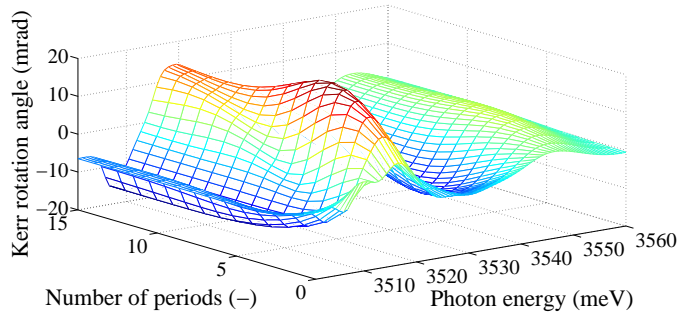}}
\caption{
Calculated Kerr rotation angle for a structure involving Ga$_{0.994}$Mn$_{0.006}$N combined with DBR layer as described in the text, plotted versus photon energy and number N of periods constituting DBR mirror.
}
\label{fig:Kerr_PE_N}
\end{figure}

In Fig. \ref{fig:Kerr_PE_N} the Kerr rotation angle is plotted as a function of the photon energy and the number N of periods constituting the DBR. The magnitude of the MOKE increases with the increasing N, saturates for about N = 4 and stays roughly unchanged after further increase of N. The saturation effect seen in Fig. \ref{fig:Kerr_PE_N} indicates a limitation of the proposed method for the MOKE enhancement, however Fig. \ref{fig:Kerr_PE_N} shows that already for a number of the DBR periods as small as 3 or 4 the MOKE is significantly enhanced with respect to "no DBR" case. This points toward ease of the sample fabrication, an important factor in a view of possible practical implementations of the proposed design.

\begin{figure}
\centering
\centerline{\includegraphics[width=0.8\columnwidth,draft=false]{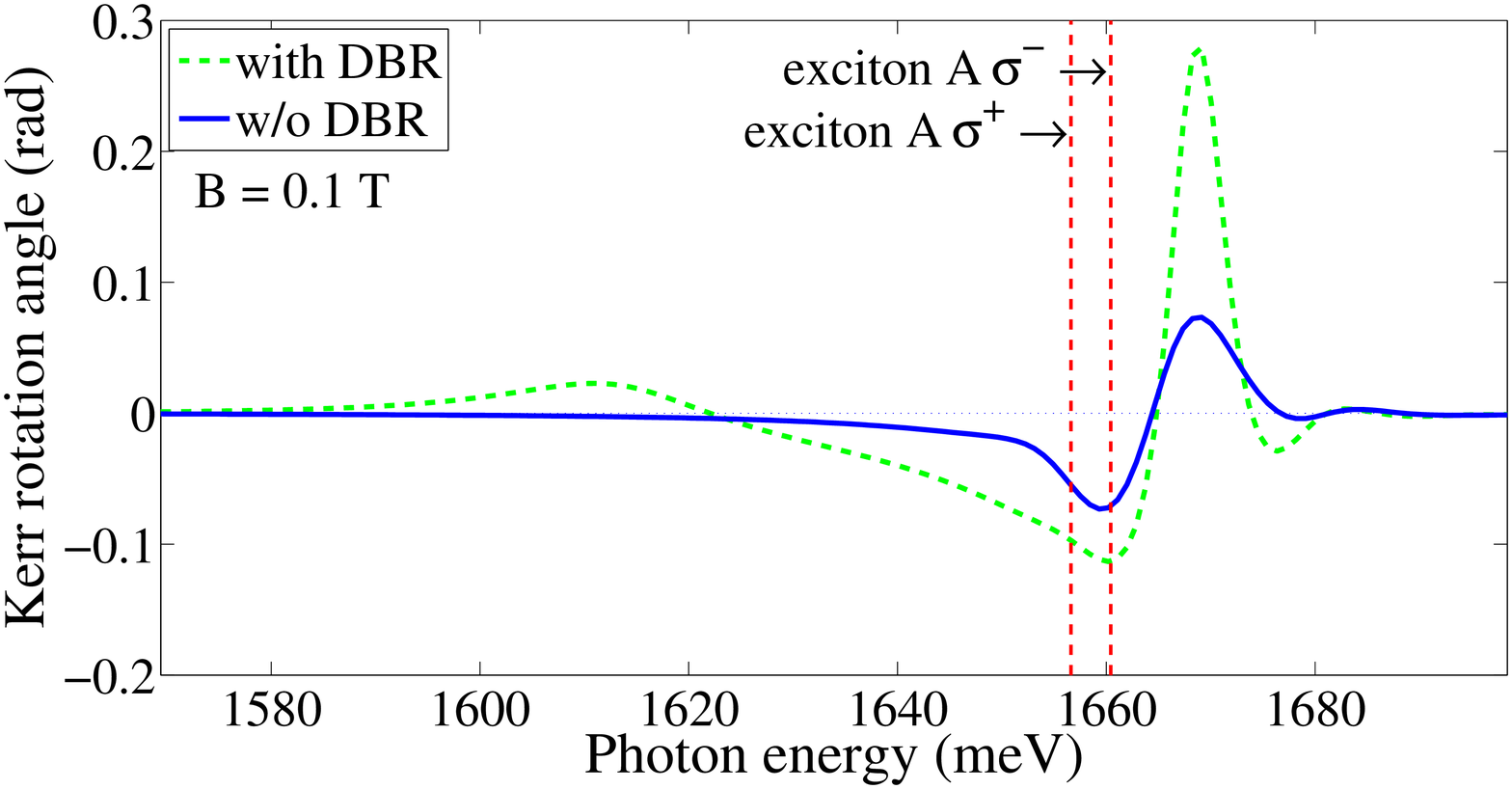}}
\caption{
Calculated Kerr rotation angle versus photon energy at B = 0.1 T and T = 4.2 K for a structure containing Cd$_{0.96}$Mn$_{0.04}$Te layer with (dashed line) and without (solid line) Cd$_{0.95}$Mg$_{0.05}$Te/Cd$_{0.7}$Mg$_{0.3}$Te Bragg mirror underneath. The position of the A exciton is indicated  for both circular polarizations of the light for a reference.
}
\label{fig:Kerr_PE_cdmnte}
\end{figure}

Finally, we apply the proposed sample design and conduct respective calculations for the structure involving zinc-blende II-VI DMS, (Cd,Mn)Te. In contrast to the wurtzite (Ga,Mn)N where three excitons contribute to the dielectric function, in the case of CdTe based alloys, a zero field splitting of the valence band is typically negligible and one deals with a single exciton contribution to the material dielectric function. We set the (Cd,Mn)Te layer thickness $d_{DMS} = 76$ nm, and Mn quantity amounting to 4\%. The DBR layer consists of Cd$_{0.95}$Mg$_{0.05}$Te and Cd$_{0.7}$Mg$_{0.3}$Te, with thicknesses of 58 nm and 70 nm, respectively. The number of periods within DBR amounts to 3. In this design, there is no buffer layer, and the substrate is made of GaAs.
As can be seen in Fig.~\ref{fig:Kerr_PE_cdmnte}, presenting the Kerr rotation angle as a function of the photon energy, the use of the Bragg mirror leads to a few fold enhancement of the MOKE, quantified by the integrated of the area under the curve. There are energy ranges where the MOKE amplitude is even more enhanced, e.g., around 1610 meV and 1630 - 1650 meV. Thus, implementation of the DBR in the case of the (Cd,Mn)Te provides a considerable MOKE increase. This indicates the utility of the proposed design in the case of both, III-V and II-VI DMS.

\section{Summary}
\label{sec:summary}
In this paper, we present the theoretical design of diluted magnetic semiconductor layer combined with the one-dimensional photonic structure (Bragg mirror) and show that  even for a relatively small number of periods and a relatively small contrast of refractive indices of layers constituting the DBR, a significant amplification of the observed magneto-optical effects (MOKE) related to DMS layer is achieved.
The presented design allows for a broadening of the range of magnetic ion concentrations within the DMS for which MOKE could be observed. In particular, it should enable detection of sizable MOKE even for boundary (very low or very high) concentrations of the magnetic dopant, where the quality of magnetooptical signal decreases. The proposed design is valid for the case of both, wurtzite III-V and zinc-blende II-VI DMS.

\ack

The work was supported by the European Research Council through the FunDMS
Advanced Grant (\#227690) within the ''Ideas'' 7th Framework Programme of
the EC, the Austrian Fonds zur F\"{o}rderung der wissenschaftlichen Forschung-FWF (P22477, P20065 and N107-NAN), and by polish NCBiR project LIDER. \\ We thank Tomasz Dietl for valuable discussions.



\end{document}